# Portability of Scientific Workflows in NGS Data Analysis: A Case Study

Christopher Schiefer[1], Marc Bux[1], Jörgen Brandt[1], Clemens Messerschmidt[1,2,3,], Knut Reinert[4], Dieter Beule[3,5], Ulf Leser[1]

[1] Humboldt-Universität zu Berlin, Knowledge Management in Bioinformatics, Berlin, Germany
[2] Core Unit Bioinformatics, Berlin Institute of Health (BIH), Berlin, Germany
[3] Charité - Universitätsmedizin Berlin, Berlin, Germany
[4] Freie Universität Berlin, Algorithms in Bioinformatics, Berlin, Germany
[5] Max Delbrück Center for Molecular Medicine, Berlin, Germany

## Abstract

The analysis of next-generation sequencing (NGS) data requires complex computational workflows consisting of dozens of autonomously developed yet interdependent processing steps. Whenever large amounts of data need to be processed, these workflows must be executed on a parallel and/or distributed systems to ensure reasonable runtime. To simplify the development and parallel execution of workflows, researchers rely on existing services such as distributed file systems, specialized workflow languages, resource managers, or workflow scheduling tools. Systems that cover some or all of these functionalities are categorized under labels like scientific workflow management systems, big data processing frameworks, or batch-queuing systems. Porting a workflow developed for a particular system on a particular hardware infrastructure to another system or to another infrastructure is non-trivial, which poses a major impediment to the scientific necessities of workflow reproducibility and workflow reusability.

In this work, we describe our efforts to port a state-of-the-art workflow for the detection of specific variants in whole-exome sequencing of mice. The workflow originally was developed in the scientific workflow system snakemake for execution on a high-performance cluster controlled by Sun Grid Engine. In the project, we ported it to the scientific workflow system SaasFee that can execute workflows on (multi-core) stand-alone servers or on clusters of arbitrary sizes using the Hadoop cluster management software. The purpose of this port was that also owners of low-cost hardware infrastructures, for which Hadoop was made for, become able to use the workflow. Although both the source and the target system are called scientific workflow systems, they differ in numerous aspects, ranging from the workflow languages to the scheduling mechanisms and the file access interfaces. These differences resulted in various problems, some expected and more unexpected, that had to be resolved before the workflow could be run with equal semantics. As a side-effect, we also report cost/runtime ratios for a state-of-the-art NGS workflow on very different hardware platforms: A comparably cheap stand-alone server (80 threads), a mid-cost, mid-sized cluster (552 threads), and a high-end HPC system (3784 threads).

## Introduction

Scientific workflows are series of programs, called tasks, connected by input-output dependencies that dictate a partial execution order [BL13]. Parallelization is essential for these workflows when the data sets to be processed are large, e.g., several terabyte in size. While mid-size data sets often only require powerful stand—alone servers, where parallelization is achieved by using concurrent threads, larger data sets require distributed infrastructures consisting of multiple nodes, each featuring multiple threads. As distributed systems bring their own complexities, such as remote file access or the need to deal with heterogeneous nodes, workflow engines typically rely on different lower-level services offering unified interfaces over distributed resources, such as file systems and resource managers. A sketch of a typical architecture of workflow systems over distributed infrastructures is shown in Figure 1. In this architecture, a workflow is specified in some workflow language and then compiled statically or dynamically into an execution plan by a workflow execution engine. The physical order in which and the node on which tasks are executed is determined by a scheduler, which uses a resource manager to access the resources of the distributed nodes. Each node has a local resource manager and a runtime environment, capable of running the tasks assigned to it by the scheduler. Tasks of a workflow exchange data through a distributed file system.

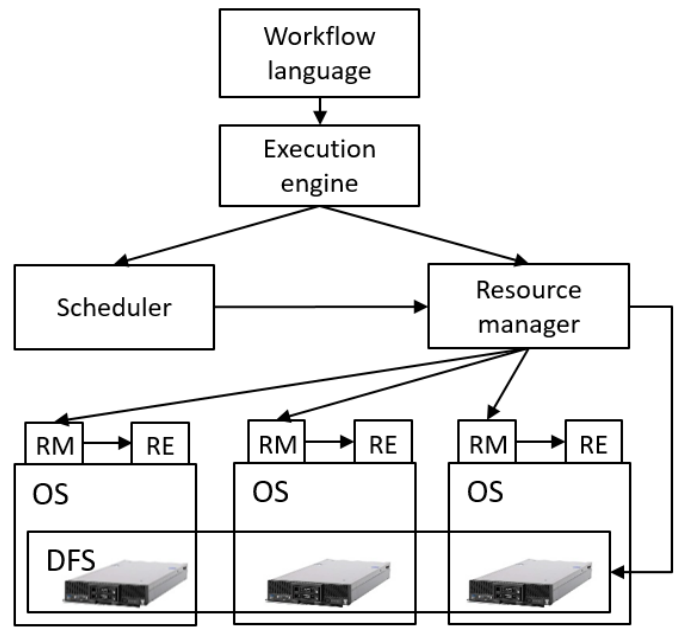

Figure 1. Architecture of a workflow system over a distributed infrastructure. RM: Resource managers; RE: Runtime engine; OS: Operating system; DFS: Distributed file system

Very few of these components are standardized, nor are their interfaces. This makes the exchange of system components difficult. For instance, most file systems support the POSIX standard for working with files, yet the popular Hadoop distributed file system (HDFS) does not [Bor07]. As another example, many batch queuing systems support the DRMAA interface for specifying jobs [TBG+12], but the recent big data processing framework Spark does not [ZCD+12]. Standardization is also lacking at the higher levels. There exists a multitude of different workflow languages with often very different semantics [BRL17]. In most existing concrete systems, no separate execution engine exists, but workflow execution, from parsing a workflow to partitioning of data and determination of physical plans, is implemented in a single, monolithic component. Resource managers often bring their own schedulers; overall, schedulers as separate, potentially interchangeable component are rare. Furthermore, concrete systems typically were designed for a certain class of distributed infrastructures. For instance, batch queuing systems, which originate in HPC systems, usually expect their execution nodes to be rather homogeneous (same memory, same CPUs etc.), whereas the recent systems big data processing frameworks usually explicitly address heterogeneity in the infrastructure in one or the other way (e.g., [WZX+16, SCN+15]). These assumptions, in turn, have consequences on the type of scheduling algorithm needed.

A scientist typically develops her workflow for the given, concrete environment available at her organization. The environment offers all or most of the described functionality by assembling specific components selected by technical administration staff. Due the lack of standardization, such a workflow typically will only run on this specific environment or on environments with only small deviations. Whenever larger changes occur, execution may break, systems may be suboptimally used, or runtimes may deteriorate drastically. For instance, workflows often explicitly encode the degree of parallelism of the workflow execution. When an infrastructure has considerably less nodes at its disposal than this parameter, runtime may suffer dramatically due to overloading of nodes; when an infrastructure has considerably more nodes at hand, most of them will stay unused, leading to ineffective execution. Due to this situation, porting a workflow from one system to another today often still is a major undertaking. This is an unfortunate situation, as it critically impedes workflow reuse and reproducibility of results [CBB+17]. Actually, we believe that most of today's large-scale computational analysis running over very large data sets are not reproducible without access to the very same infrastructure the original analysis was developed for[1].

In this paper, we report on the process of porting a workflow for finding specific genomic variants in certain mice. The workflow originally was written with snakemake [KR12] for a high-end cluster controlled by the Sun Grid Engine (SGE – now Oracle Grid Engine[2]) and featuring a high-performance parallel file system. As many labs have no access to such costly infrastructures and such infrastructures furthermore are not necessary for moderately large data sets, we aimed at porting it to the scientific workflow engine SaasFee [BBL+15]. SaasFee can execute workflows on any cluster running the popular resource manager Yarn, a resource manager specifically developed for large clusters of inexpensive hardware with standard network facilities. SaasFee also supports stand-alone servers, which are much easier to administer, cheaper to acquire, often fast enough for smaller data sets and therefore very attractive for smaller labs with small or moderate data production. During this port, we had to (1) translate the workflow from snakemake, an extension of Python, to Cuneiform [BBL15], the workflow language of SaasFee, (2) change the workflow to adapt to the different infrastructures, in particular the different file access methods, (3) change the workflow to adapt to the very different scheduling policies of SGE and SaasFee, (4) change SaasFee's execution engine Hi-Way [BBW+17] to cope with certain requirements of the workflow, and (5) perform numerous experiments to detect and to solve subtle issues created by differences between source and target system. After roughly three net person months of work, we eventually were able to run the workflow simultaneously on the source and the target system.

The aim of this paper is to inform other scientists facing the challenge of porting their workflow across heterogeneous systems about the problems and pitfalls we encountered. Clearly, this is not a systematic study on workflow portability, which would require investigating a multitude of source / target pairs, but a detailed case report on a specific port.

As a side-effect, we eventually also were able to compare the runtimes of the workflow on different systems over the same input, and to put them into perspective to the acquisition costs of the systems. As SaasFee can also run on cloud-based virtualized infrastructures, we were also able to estimate runtime and costs when running the workflow on rented virtual machines from Amazon's EC2.

## Methods and Data

### Overview of the Workflow

The workflow described here has been developed as part of a project for characterizing changes of Diffuse Large B-Cell Lymphoma (DLBCL) in mice. The workflow itself is only concerned with exome sequencing. DNA samples from mice suffering from DLBCL and healthy control tissue were whole-exome sequenced (WES) using an Illumina HiSeq machine[3]. Enrichment of exonic sequences was performed using Agilent SureSelect XT Mouse AllExon. The exons of a gene make up the mature mRNA, which is the blueprint for the final protein. The prime interest of the project was to detect somatic variations in the genome between tumor and healthy tissue, as these can offer insights into the functioning of a tumor. In the original experiment, 27 samples from DLBCL and two controls were sequenced and their 27*2=54 pair-wise genetic differences were computed.

---

[1] We admit that we have no experimental evidence for this claim.
[2] See https://www.oracle.com/technetwork/oem/grid-engine-166852.html
[3] See https://www.illumina.com/systems/sequencing-platforms.html

In this work, we only focus on the core workflow, starting from the set of reads created by the sequencer and ending with sets of genetic differences[4]. Furthermore, for simplicity we here only consider single nucleotide variants, although other variations are equally important. An overview of this workflow is shown in Figure 2. It takes as input the read sets of each sequenced sample and first uses BWA-MEM [Li13] to align the individual reads to a mouse reference genome (MGSCv37 - mm9). After creating an index of the aligned sequences, each genome from a tumor is paired with the genome of a healthy control to find somatic variants using MuTect [CLC+13]. As the last step, the workflow removes likely non-somatic variants and compresses the output files.

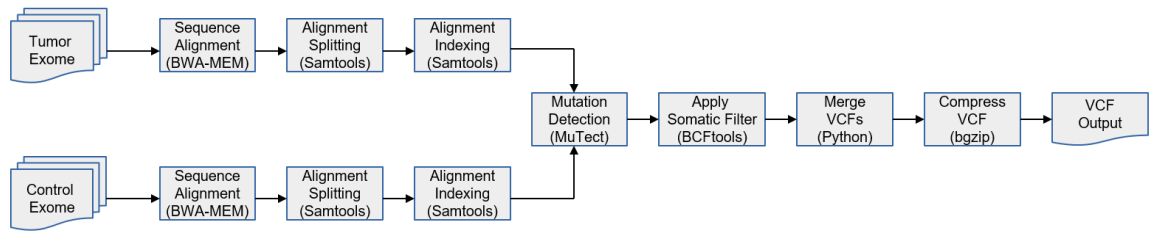

Figure 2. Workflow for finding variations in whole-exomes of two samples. Tools are described in more detail in the text and in Table 1.

For porting this workflow from one scientific workflow system to another, especially the possibilities for parallelization are important. For describing these, it is instrumental to divide the workflow into three phases:

- **Alignment phase**: In the alignment phase, all samples, i.e., read sets, can be aligned independently from each other and in parallel. Furthermore, each read can be aligned independently of the other reads of a read set (up to a certain point of the analysis), offering further parallelization options within the aligner itself. In BWA-MEM, the latter can be configured by a parameter dictating the number of threads to be used during alignment. In contrast, parallelization at the sample level must be programmed in the workflow itself.

- **Variant calling phase:** In the variant calling phase, pairs of aligned samples are analyzed by the tool MuTect to identify point-wise differences. MuTect uses only a single thread; however, the genomes of the samples can be partitioned prior to giving them to MuTect. Thus, this process can be parallelized by starting different MuTect instances, each on a different, non-overlapping region of a genome.

- **Post-Processing phase:** In this phase, the sets of variants created by the region- and pair-wise comparison of tumors to samples are merged, non-somatic variations are removed, and the overall output is compressed. The first two steps of this phase can be parallelized at the level of the regions defined for the previous phase; in contrast, the compression requires the genome-wide lists as input.

An overview of the resource characteristics of the different tasks in the workflow is shown in Table 1. The two most resource-hungry and hence most important steps are BWA-MEM and MuTect:

- BWA-MEM loads a reference genome and a read set into main memory and scales the alignment across all threads allowed by the command-line parameter. Due to the algorithmic challenges of read alignment, it is a CPU- and memory intensive program, whereas IO plays a minor role. Network traffic is mainly generated at the beginning when the FASTQ files are distributed throughout the cluster nodes. On a single node with a fixed size of main memory and a fixed number of threads, one can start multiple instances of BWA-MEM, but must consider that each instance requires a certain amount of memory. If more instances are started than memory is available, the system will start to spill memory on disk, leading to terrible runtime delays; if more instances are started than threads are available, runtime will slowly degrade, as instances have to wait for each other.

---
[4] The original workflow implementation was performed in 2016 and the port was accomplished in 2017. Some of the tools used today have other configurations.

| Tool | Task | Multi-Threaded? | Memory requirements? | Network requirements? |
|---|---|---|---|---|
| TRIMADAP-MT[5] | Sequence trimming | Yes, configurable | Low (streaming) | Low at start to access read files and at end for writing results. Tools communicate via memory pipes. |
| BWA-MEM [Li13] | Read alignment | Yes, configurable | High (8-14 GB) | |
| SAMBLASTER [GH14] | Mark read duplicates | No | Low (streaming) | |
| SAMTOOLS [LHW+09] | Alignment splitting, and indexing | Yes, configurable | Low (streaming) | |
| MuTect [CLC+13] | Mutation detection | No | Medium (3-10 GB) | High |
| BCFTOOLS[6] | filtering somatic mutations | Yes, configurable | Very low | Very low |
| Script (python) | Merge VCF files | No | Very low | Very low |
| bgzip | Compress VCF files | No | Very low | Very low |

Table 1. Overview of important characteristics of the tools in the workflow.

- In contrast, MuTect, when run on smaller regions, is a more network-bound process. For each region in each combination of control and tumor exomes, two alignments have to be sent to the respective MuTect task. If smaller regions are chosen, more MuTect instances can be run in parallel, but also more network traffic is incurred. If larger regions are chosen, less instances are started and the runtime of a single instance starts to matter. Additionally, each process requires the full reference dictionary in the main memory (2.6 GB), which must be considered when determining the number of MuTect instances to start on a single node.

The properties have obvious consequences regarding the infrastructure at hand. For instance, BWA-MEM benefits from strong nodes with many threads and a large main memory, whereas MuTect is sensitive to network bandwidth and data shipment protocol. These consequences will be discussed in more details below.

| Name (abbreviation) | Stand-Alone (SA) | Yarn-Cluster (YC) | HPC Cluster (HPC) |
|---|---|---|---|
| Type | monolithic | cluster | cluster |
| Controlled by | SaasFee | SaasFee | Snakemake-SGE |
| Compute nodes | 1 | 23 | 111 |
| Total number of threads | 80 | 552 | 3784 |
| Memory / node | 512 GB | 24 / 36 GB | 128 / 188 / 500 / 1000 GB |
| Mem. / thread (average) | 6.4 GB | 1.25 GB | 11.7 GB |
| Network type | None | Ethernet | Ethernet |
| Bandwidth | Na | 10 GBit | 64 GBit |

Table 2. Overview of the hardware infrastructure of the source system (third column) and of the target systems (first and second column).

---

[5]https://github.com/lh3/trimadap
[6]http://samtools.github.io/bcftools/bcftools.html

**Source and target system**

In this project, we ported the workflow just described from a source system, henceforth called snakemake-SGE, to a target system, henceforth called SaasFee. These exhibit distinct properties that will be described in the following paragraphs. The main aim of the port was to make the workflow, developed for an expensive high-end system, also available for users having only significantly less efficient and less costly hardware infrastructure. An overview on the different infrastructures used in the experiments is shown in Table 2.

- **Snakemake-SGE**. The workflow originally was implemented in Python using the scientific workflow system snakemake. A workflow is defined by a set of rules and each rule specifies a set of input- and output-files as well as instructions (e.g. in Python or bash) for transforming the input files into the output. Upon execution, Snakemake creates a DAG with all tasks and automatically determines which tasks can be run in parallel. This DAG is transferred to a Sun Grid Engine through the DRMAA interface. SGE schedules the tasks on all available nodes with the goal to achieve uniform load, taking resource constraints, such as limited memory, into account. The location of input (and output) files is not considered for scheduling, because the engine assumes a fast network and fast shared file system. The specific system described here uses the General Parallel File System (GPFS).

- **SaasFee**. SaasFee is a scientific workflow engine consisting of the workflow language Cuneiform and the execution engine HiWay. Cuneiform is a functional workflow language in which tasks are encapsulated as functions. These can be written in multiple programming languages, they can be chained, and they can be used in second-order functions, like MAP, CROSS, or REDUCE [BBL15]. A Cuneiform program is compiled into a logical workflow execution plan. This plan can be executed either on a stand-alone server using only local main memory and the local file system, or it can be passed to the workflow engine HiWay. HiWay coordinates the execution of a workflow plan on a cluster of nodes managed by Apache Yarn, using HDFS as distributed file system[7]. Whatever mode is used, a runtime environment must be installed on all participating nodes to ensure that assigned tasks are executable. As HDFS is not POSIX compliant yet tasks typically access their input file through a POSIX interface, HiWay also takes over responsibility to stage-out input data from HDFS to the node chosen for executing a task, and to stage-in output files into HDFS to make them available on the entire cluster.

**Original implementation**

The original implementation of the workflow targeted the snakemake-SGE stack on the HPC cluster (see Table 2). It first paired each control exome with each tumor exome, yielding 54 pairs in total. For each pair both FASTQ files are first aligned with BWA-MEM, making up 108 alignment tasks. Note that most alignments are redundant because BWA-MEM processes each tumor exome two times, and each control exome even 27 times, once for each tumor input file. Such inefficiencies are not rare on systems with large compute power if they are faster to develop and the additional compute time is not considered critical. Through manual configuration of the workflow, SGE is instructed to use 54 nodes (roughly half of the HPC cluster) and to process on each node two instances of TRIMADAP, BWA-MEM, SAMBLASTER and SAMTOOLS with up to 24 threads. Within this sub-workflow, all tools communicate via pipes, avoiding any IO. Each instance of this sub-workflow splits the resulting alignment into 467 genome regions to allow parallel processing with MuTect in the next phase. One MuTecT instance is started for every region in each of the pairs, resulting in 467 * 54 = 25.228 tasks. These variant-calling tasks are distributed across all available nodes on the cluster. In the end, all 467 result files of each pair are combined into a single compressed VCF output file.

## Results

We describe our experiences in porting a workflow for NGS data analysis from a system based on snakemake / SGE / GPFS on a HPC cluster with a fast network to a system based on SaasFee / Yarn /

---
[7] See https://hadoop.apache.org/

HDFS on a stand-alone server or a modestly sized cluster with a moderately fast network. We structure our report along the components of a scientific workflow system as sketched in Figure 1 – starting from the workflow language to the resource management / scheduling and eventually to the file system. We also describe bugs and inefficiencies we encountered and corrected during the port, both in the workflow itself and within the systems used.

**From Snakemake to Cuneiform**

The original implementation was based on snakemake, an extension of Python for defining and executing scientific workflows. The target system was programmed in Cuneiform, a functional scientific workflow language. The two languages follow a very different approach to workflow specification – while snakemake is declarative, defining tasks and dependencies in terms of rules that together and implicitly span the workflow DAG, Cuneiform is a functional language, where dependencies are explicitly defined by the developer.

Porting between these languages thus had two aspects. First, the basic tasks had to be extracted from the snakemake program and wrapped in Cuneiform functions. As Cuneiform supports multiple programming languages in its functions including bash and Python, this part was rather straightforward. The only changes we applied were the "uncovering" of certain parameters, such as the number of regions to split in the aligned reads. These were hard-coded within scripts in snakemake, whereas we had to expose them as Cuneiform parameters in the target system to improve control for execution on different target infrastructures (see below).

Specifying the structure of the workflow required considerably more attention. We reengineered the call structure from the snakemake program by manually comparing rule results and preconditions, and translated them into Cuneiform constructs. Special attention was necessary for loops. These are (implicitly) encoded in snakemake as rules matching to certain file names, which we had to translate into separate Cuneiform functions. These functions are concerned with managing the relationships between the different tasks, e.g. to specify whether lists of arguments are to be combined as dot product or as cross product.

**Scheduling and resource management**

Scheduling workflows on parallel and/or distributed infrastructures is an intricate problem. Optimal schedules depend on the available number of nodes, the tasks to be scheduled, the speed of the underlying network and file system, the resources of each node (especially memory and number of hardware threads), the robustness of the system in terms of failure recovery, and many more factors [LPVM15]. The tasks to be scheduled, in turn, are designed by the workflow developer, which involves decisions regarding their granularity. In principle, having more (independent) tasks creates more degrees-of-freedom for the scheduler, but also poses a more complex problem to the scheduler and leads to more communication between tasks, which may be expensive if the file system is involved. The consequences of these trade-offs differ much between different systems and different infrastructures, and it is largely the responsibility of the developer to find a suitable configuration, where suitable can mean anything from "satisfying" to "optimal".

In principle, the two systems studied here do not differ much in their architecture with regard to the execution engine. In the source system, snakemake creates a workflow dependency graph that is handed over to the SGE for scheduling. In the target system, Cuneiform programs are compiled to a workflow dependency graph[8] that either is scheduled locally (SA system) or is handed over to HiWay for scheduling (YC system). However, notable differences exist in the lower levels of the workflow environment (see Figure 1), which has severe implications on scheduling and resource management as well. First, the HPC source system uses a shared parallel file system over a fast network. The scheduler thus simply ignores the location of input files when assigning tasks to nodes. In addition, the original workflow implementation tended to ignore file access and network issues, as it was considered fast enough. In contrast, the YC target system has an only moderately powerful network. Furthermore, input and output files must be staged-in and staged-out from / to HDFS, creating additional IO when tasks access data that is not available locally. To minimize such cases, the HiWay scheduler is placement-aware, i.e., it tries to place tasks at those machines where their input data already (partly or entirely) resides, trying to avoid the detour over HDFS. Second, the available main memory differs much between the three systems we studied. Nodes in HPC (source) have large memories, allowing many task instances to run on the same node. Also SA (target) has a large memory compared to the number of threads. In contrast, the nodes in YC (target) have smaller memories, which can quickly induce paging when too many instances are started in parallel.

---

[8] Note the in SaasFee the interface between the execution engine and the Cuneiform runtime engine actually is dynamic, also allowing for dynamic workflow graphs [BBW+17]. However, as this functionality is not important for the workflow described here, we omit this detail.

We describe the consequences of these differences regarding the two main components, BWA-MEM and MuTect:

- **Adjusting BWA-MEM**. This tool is convenient to handle as one can increase the number of workers for each task without requiring more memory, because BWA-MEM always requires the same memory to load the entire reference genome (in indexed form). This makes the ratio between number of threads and available RAM adaptable. We configured the target workflow on SA to start 30 alignment tasks in parallel with three threads each. This leads to an almost 100% usage of CPU resources without exceeding the main memory. In contrast, in YC we could start only three or four instances per node, as these already completely exhausted memory resources. This implies that the result files are distributed over the entire cluster and must be shipped later for the MuTect tasks. The HPC system has a convenient surplus of RAM for this task, with each single thread having more memory available than a BWA-MEM instance actually consumes. One could comfortably run more than 3000 parallel alignment tasks on this system, significantly outnumbering the maximum of 108 tasks needed for this workflow.

- **Adjusting MuTect**. MuTect, in the version we used for our workflow, was a single-threaded tool. Thus, the number of instances to start is determined by the number of disjoint genomic regions created in the previous phase. Independent of the size of the region to work on, each MuTect instance requires an almost identical amount of main memory (roughly 3 GB) for loading the reference genome. Therefore, it depends on the specific hardware infrastructure into how many parts a sample should be split. With only 1-2 GB RAM available on average per thread, YC can only utilize a fraction of its CPU capacity: Execution of MuTect here is memory-bound, not CPU-bound. Starting more instances on a node will not lead to more parallelization due to the lack of free memory. In contrast, SA can start 80 MuTect instances for its 80 threads without running into memory shortage. As all IO on SA is local, a low number of splits could be attractive in this setting, as it leads to a low number of tasks, making scheduling and resource management easier. This also holds for the HPC system, as each of its nodes has enough memory to start as many MuTect instances as threads are available.

The main configuration parameter to adapt the workflow to different hardware infrastructures thus is the number of splits of an aligned sample pair. In the source system, the number of splits was fixed to 467. As there is a trade-of between network traffic, complexity of scheduling, and load per node regarding the number of splits, in the target system we experimented with three different configurations: 22, 467, and 2863 (see discussion).

**Execution and file access**

Each workflow system must ensure that each node it assigns a task for execution can actually execute this task. This implies that the code must be executable, which requires properly compiled binaries, that libraries a task depends on are available in the right version at the expected position, and that the file system structure is as assumed by the task. The two most prominent ways of ensuring this are "virtualization" and "installation". The latter is the more conventional way and requires installation of a proper runtime environment (i.e., all task binaries and its dependencies) in the operating system on every node participating in the execution. The former today is the preferred way. Here, tasks and their dependencies are installed once in a virtual environment, e.g. a container or a virtual machine, which is then installed on every node. As at the time of this study container technology was in its infancy and virtual machines were considered performance-harming, the source system applied "installation". Porting the **runtime environment** therefore required recompiling all tools on the specific target platforms and gathering platform-specific versions of all its dependencies. In Hi-Way, proper environments are build by installing a "raw" virtual machine for every task execution on every node, which first executes a pre-prepared script that installs the runtime environment within the VM. Porting the runtime environment therefore required a time-consuming programming of this script for installing roughly a dozen of tools in the VM.

Another difference between the two systems is **file access**. In the HPC system, all tasks access their files over a shared parallel file system. Thus, they are not aware of and need not bother with the location of their input or output files. As the underlying network is fast, optimizations of IO, by bringing tasks closer (network-wise) to their data was not implemented. In contrast, the target system on YC uses HDFS for file access. HDFS also is a shared parallel file system but its API is not POSIX-compliant, which means that task not implemented on the HDFS API cannot use it. This, in turn, implies that each task has to be wrapped in a script that first reads all input data for a task from HDFS, places it in the local file system, configures the task to use this data, reads all its local output files, and eventually places them in HDFS to make them available for future tasks. This procedure is transparent for the workflow developer, as HiWay automatically applies it for every task invocation. Nevertheless, it implies that workflows in SaasFee are very sensitive to large amounts of IO, making certain adaptations of the workflow necessary to achieve "fast enough" runtime on YC. The situation is much more comfortable on the SA system, as this is a stand-alone server where all IO is local by design (of course, one should avoid the usage of remotely mounted file systems). On SA, Cuneiform creates a separate working directory for each task it starts. This directory is provided a copy of all input files and used for all intermediate and result files. Directories are not shared between tasks, so that all task-specific files are copied with each invocation. This also creates duplicate IO, but largely improves debugging and failure handling.

| System | SA | YC | HPC |
|---|---|---|---|
| Estimate of acquisition costs | ~11.000 € | ~100.000 € | ~800.000 € |
| Workflow runtime (no caching) | 65.13h | 11.13h | 1.14h |
| Workflow runtime (caching) | 24h | 7.6h | - |
| Theoretical throughput per year (with caching) | 365 runs | 1153 runs | 7684 runs |
| Effectiveness (Throughput per Euro) | 0,033 | 0,012 | 0,019 |

Table 3: Cost, runtime, and effectiveness (as throughput per Euro) for three different systems running the same workflow over the same input and producing the same results. All runs were using 467 splits of each sample pair. For the effectiveness computation on HPC, we used only 50% of the HPC acquisition costs, as the workflow uses only 50% of the nodes of this system.

**Cost/runtime Ratios on Different Systems**

The main aim of this study was to make the workflow available also for users and institutions without access to a very powerful compute infrastructure. After porting the workflow, we were able to execute it on three different systems, namely the source system HPC and the two different target systems SA and YC. Table 3 shows, for each system, the acquisition costs, the runtimes (with/out caching on the target system), the theoretical throughput per year, and the effectiveness, computed as throughput per Euro. Not surprisingly, the most powerful system (HPC) clearly is the fastest. It however, has a much worse effectiveness than the stand-alone server. Its effectiveness is roughly the same as that of the YC system. The latter may seem surprising, as the acquisition costs per node are considerably lower. However, we have seen that runtimes on YC suffer from insufficient memory per node and from the need of performing much additional IO through the lack of a POSIX-compatible distributed file system. In this sense, it is interesting that the improved scheduling of Hi-Way seems to be able to – at least partly – compensate for these disadvantages.

As SaasFee is also able to run workflows on rented nodes in commercial clouds, we also performed experiments to estimate cost and effectiveness for such an installation. Note that the workflow only analyses mice genomes, which means that data privacy restrictions would not apply. Using 16 nodes of type r3.4xlarge, each having 256 threads and 122 GB, we estimated that the workflow execution would take ~14h and would incur a cost of ~500 Euros. Using this setup, the throughout per year would be ~625 at a total cost of ~313.000 Euros, yielding an effectiveness value of 0,002. As one would expect, renting resources is much less cost-effectiveness than buying hardware in cases where hardware utilization is maximal. If we assume that, for instance, only 50 such experiments per year are performed, we get an effectiveness of 0,0045 for SA, 0,0005 for YC, 0,0001 for HPC, and 0,002 for EC2. At such low system utilization rates[9], HPC becomes the least cost effective system; EC2 still is considerable less effective than a stand-alone cluster, but would perform every run roughly twice as fast; and YC becomes the second best effective system, also running roughly three times faster then SA.

## Discussion

### Bugs and Inefficiencies

During our port, we encountered a number of bugs and/or inefficiencies both in the source and in the target system. We describe these in the following. The first two are inefficiencies in the original workflow implementation which we corrected, as the target systems are much less powerful than the source system, which greatly increases the pain accompanied with such inefficiencies. The other two issues are bugs we found in the SaasFee implementation, which here, for the first time, was confronted with workflow execution plans consisting of several thousands of physical tasks.

- **Adjusting the transition from BWA-MEM to MuTect.** Any MuTect instance in the workflow analyzes only a small region of the alignment of a sample pair. However, "splitting" an alignment in regions can be implemented in two ways. In the original implementation, the output of BWA-MEM was not split, but sent entirely over the network to every MuTect instance. These instances were configured to process only a specific region by command-line parameters. This created a lot of unnecessary IO, as most of the transferred data is not used by the receiving MuTect instances. Therefore, we adapted the workflow such that it physically splits the output of BWA-MEM into separate files, such that each file represents a region that an individual MuTect instance should analyze. Only these regions were subsequently sent over the network.

- **Redundant BWA-MEM instances.** In the original workflow, a total of 108 instances of BWA-MEM are started: Two for every sample pair. However, only 29 alignments are actually necessary, because a sample always creates the same alignment, irrespective of its partner in a sample pair. Fortunately, SaasFee has a "caching" feature, which caches the result of task invocations and reduces the cached results in case the very same task is started again. This feature can be switched of, as it only works properly for strictly deterministic tasks. We run the experiments both with caching activated and deactivated for increased comparability (see Table 3).

- **SaasFee with many tasks**. Running the workflow with Cuneiform revealed deficiencies in its implementation. Especially the large number of physical tasks generated by the workflow caused problems for task management. Beforehand, tasks and their designated folders were organized using a hash function. For smaller workflows, this function only very rarely created hash collisions. However, hash collisions became rather frequent for the workflow studied here, causing problems for the implementation. Additionally, the interpreter originally traversed the workflow graph with linear complexity for each update, leading to a complexity that is quadratic in the workflow size. This notably increased the total run time for a single run of the workflow. Both problems were solved with a new version of Cuneiform. The high number of tasks also caused initial problems in Hi-Way as the default configuration assigned too little memory to the master node to store and process its log. This issue was solved by assigning more memory to the process.

---
[9] We believe 50 is not an unrealistic number for computational infrastructures that are not run within dedicated, department-spanning service infrastructures.

- **Task memory assignments in SaasFee**. We experienced various pitfalls within SaasFee regarding memory assignment on both target systems. On the SA system, memory and thread assignment is performed directly by the Cuneiform runtime engine, which (at that time) had a single parameter for specifying the maximum number of concurrent tasks within a workflow execution. However, the two main phases of the workflow exhibit different memory requirement. As the BWA-MEM phase has high memory requirements, only 30 instances can be started concurrently, which required to set the parameter to this value to avoid overload. Accordingly, Cuneiform also in the second phase started only 30 MuTect instances in parallel, although MuTect requires less memory. This behavior left considerable compute resources unused and significantly increased the total runtime. On the YC system, workflows are scheduled by HiWay using the Yarn framework. To perform sensible resource management, HiWay (in contrast to Cuneiform) has a task-specific parameter to configure the memory given to each instance of this task. However, MuTect has different (not drastically, but notable in the range of 30%) RAM demands for different input sizes. Furthermore, instrumenting the code to perform reliable memory measurements on different platforms (SA, YC) with different runtime engines and different CPUs turned out to be difficult. As memory is scarce on YC, we had to perform extensive testing to find the smallest such parameter suitable for all nodes on the platforms. Note that snakemake has a very similar setup in terms of memory assignment, but as the HPC system has considerably more memory per node than SA or YC, the respective parameters were simply set to a very conservative value using educated guessing. We did not measure the amount of memory thereby claimed but unused, which often is considerable when users are asked for estimates [DK14].

**More or less MucTect splits**

As described above, the number of regions and thus of variant calling tasks is variable. In the original workflow, the value was set to 467, producing 467 MuTect tasks per pair. In the target system, we observed trade-offs on YC between this number of the amount of network traffic, IO, and instances per nodes. To explore this trade-off, we repeated the measurement with two other values, once with much less regions (22) and once with many more (2863). Results are shown in Table 4 (caching enabled). Having many more regions considerable slowed down workflow execution, which we attribute to the more complex scheduling (467 regions generate 25.218 MuTect tasks, while 2863 generate 154.602 tasks) and the many more stage-in / stage-out processes necessary for using HDFS. In contrast, using less splits reduced the runtime by ~17%. Figures SM1 and SM2 illustrate the system load on YC depending on the number of MuTect tasks.

| Number of regions | 22 | 467 | 2863 |
|---|---|---|---|
| Runtime | 6.35h | 7.6h | 15.2h |

Table 4: Runtime on YC with varying numbers of regions and hence MuTect instances.

**Acquisition costs and total cost of ownerships**

Our analysis makes several simplifying assumptions regarding the acquisition costs of the infrastructures used. For instance, HPC systems require dedicated and powerful cooling that often requires specialized buildings; such costs have not been included here. They are, however, also often shared between groups, reducing the costs per unit. HPC systems also are maintenance heavy, required a significantly higher number of technical and administrative staff. In contrast, a stand-alone server can easily be placed in an ordinary office and can be administered and programmed by the same person. Mid-size clusters like YC are somewhat in-between, typically requiring only off-the-shelve cooling via air conditioning. Maintenance costs are not as low as in the SA setting, and the low cost components used typically result in more frequent hardware failures than in a HPC setting.

We are not aware of any comprehensive study on total cost of ownership for different computational infrastructures for genome data analysis. There are studies on the cost of genome sequencing in general [CKK+18, PFvdS17], but these focus on the cost of sequencing, not on the cost of data analysis. In their meta study, [SBTW18] found that many studies on the cost of sequencing fall short in considering all cost aspects or in being transparent regarding the aspects being included in the respective cost analysis. [MLL+16] were the first to bring up the issue of the increasing costs of data analysis in genome research. [SCFB15] analysed in detail the cost of variant calling from ~2500 WGS samples in a cloud setup but did not compare infrastructures.

**Related Work**

The advent of the "big data" area brought a flood of new systems that can be considered as scientific workflow engines [BL13, LPVM15,]. Quite a few of these were specifically developed for the analysis of large biomedical datasets in general and NGS data in particular. Examples include Galaxy [GNT+10], NextFlow [DTC+17], OpenAlea [PFV+15], and snakemake [KR12], handled in depth here. A short survey on NGS pipelines and their implementations with workflow engines can be found in [Lei17]; [SBR+15] contains a series of descriptions of practical installations in Europe.

Most of these systems follow the black-box approach also underlying snakemake and SaasFee, i.e., the workflow engine considers individual tasks as given and immutable. In these systems, scheduling becomes the prime component to influence throughput and resource consumption. Another line of research studies how individual tasks themselves can be implemented in a manner that allows for better scalability on distributed infrastructures, e.g. [RDM+12]. Here, performance engineering typically is achieved at a lower level by porting individual algorithms to APIs created for supporting distributed applications, such as MapReduce (at a logically rather high level, [ZLJ+13]), or MPI (at a logically very low level, [GGL+99]).

One aim of the present study was to support reproducibility of workflow analysis by becoming less dependent on a particular hardware infrastructure. [CBB+17] gives a general survey of reproducibility issues in the Life Sciences. Another aim was to illustrate potential users of the workflow the implications of choosing a hardware platform in terms of throughput and acquisition costs. [SL10] discussed further criteria for selecting compute platforms for big data applications. Much of the problems we encountered can be attributed to the scheduling components of the systems under study, which, in one or the other way, have to cope with the difficulties to estimate and exploit precise estimates of resource requirements of the individual workflow tasks. [WGL18] describes the state-of-the-art in predictive performance modelling in workflow-like systems using a black-box approach.

Despite the many works advocating reproducibility in large-scale data analysis [KKLS17, NT14, SSM18, MWHW16, MDBL16], there are surprisingly few studies on the difficulties of porting a concrete workflow between concrete systems. [SSVS15] compared the implementation of Crossbow, an application for read alignment and SNP calling, on a HPC system with that on a Hadoop-based cluster. [OST+19] describes porting a workflow in molecular dynamics between two supercomputers having different topologies and architectures. [CMX+16] ported a workflow for analyzing whole-exome sequencing data from a HPC system to the Microsoft Azure cloud.

## Conclusions

We described a series of problems we encountered when porting a state-of-the-art workflow (in 2016) for NGS data analysis between two scientific workflow systems with rather different technical features and running on three very different hardware infrastructures. Overcoming these problems was feasible, but required almost three person months of work, although we had the favorable circumstances that experts for the workflow and for the source and the target systems were readily available and involved in the project. The main reasons for the problems were (1) implicit assumptions on schedulers and infrastructure leading to certain design decisions in the original workflow that had to be uncovered and translated into the target system in a laborious trial-and-error process, (2), the different approaches to scheduling and file exchange in the different systems that needed to be overcome by changing the workflow design and its configuration, and (3) bugs in the target systems that were not encountered in previous experiments.

The project described in this work mostly was carried out in 2017, and the original workflow was implemented in 2016. Probably, a number of design decisions would be taken differently today. For instance, the use of containerized tool implementations make the installation of a runtime environment considerably simpler than four years ago. The increasing popularity of CWL as a standard workflow language might solve the problem of porting between languages with different semantics. However, we note that both of these issues were real, yet not responsible for most of the porting effort. We believe that the main sources of problems, i.e., heterogeneous workflow components beneath the language, implicit assumptions leading to hard-coded design decisions in the workflows, and bugs in the workflow systems, which mostly are research prototype and not mature production-ready systems, today are not less severe.

## Achnowledgements

We thank Manuel Holtgrewe from CUBI/BIH for helping with the original workflow. We acknowledge financial support through the Deutsche Forschungsgemeinschaft, Research Training Group 1651 SOAMED, the European Framework 7 programme BioBankCloud, and an Amazon research.## References

[BBL+15]   Bux, M., Brandt, J., Lipka, C., Hakimzadeh, K., Dowling, J. and Leser, U. (2015). "SAASFEE: Scalable Scientific Workflow Execution Engine". PVLDB, Hawaii, USA.

[BBL15]    Brandt, J., Bux, M. and Leser, U. (2015). "Cuneiform -- A Functional Language for Large Scale Scientific Data Analysis". EDBT Workshop Beyond Map&Reduce, Bruessles, Belgium.

[BBW+17]   Bux, M., Brandt, J., Witt, C., Dowling, J. and Leser, U. (2017). "Hi-WAY: Execution of Scientific Workflows on Hadoop YARN". Int. Conf. on Extending Database Technology, Venice, Italy.

[BL13]     Bux, M. and Leser, U. (2013). "Parallelization in Scientific Workflow Management Systems". CoRR/abs:1303.7195.

[Bor07]    Borthakur, D. (2007). "The hadoop distributed file system: Architecture and design". Hadoop Project Website.

[BRL17]    Brandt, J., Reisig, W. and Leser, U. (2017). "Computation Semantics of the Functional Scientific Workflow Language Cuneiform." Journal of Functional Programming 27(e22).

[CBB+17]   Cohen-Boulakia, S., Belhajjame, K., Collin, O., Chopard, J., Froidevaux, C., Gaignard, A., Hinsen, K., Larmande, P., Le Bras, Y., Lemoine, F., et al. (2017). "Scientific workflows for computational reproducibility in the life sciences: Status, challenges and opportunities." Future Generation Computer Systems 75: 284-298.

[CKK+18]   Christensen, Kurt D., Kathryn A. Phillips, Robert C. Green, and Dmitry Dukhovny. "Cost analyses of genomic sequencing: lessons learned from the MedSeq Project." Value in Health 21(9):1054-1061.

[CLC+13]   Cibulskis, K., Lawrence, M. S., Carter, S. L., Sivachenko, A., Jaffe, D., Sougnez, C., et al. (2013). "Sensitive detection of somatic point mutations in impure and heterogeneous cancer samples". Nature Biotechnology, 31(3), 213.

[CMX+16]   Cała, J., Marei, E., Xu, Y., Takeda, K., and Missier, P. (2016). "Scalable and efficient whole-exome data processing using workflows on the cloud". Future Generation Computer Systems, 65, 153-168.

| | |
|---|---|
| [DK14] | Delimitrou, C., and Kozyrakis, C. (2014). "Quasar: resource-efficient and QoS-aware cluster management". ASPLOS, pp. 127–144. |
| [DTC+17] | Di Tommaso, P., Chatzou, M., Floden, E. W., Barja, P. P., Palumbo, E. and Notredame, C. (2017). "Nextflow enables reproducible computational workflows." Nat Biotechnol 35(4): 316-319. |
| [GGL+99] | Gropp, W., Gropp, W. D., Lusk, E., Lusk, A. D. F. E. E., and Skjellum, A. (1999). "Using MPI: portable parallel programming with the message-passing interface". MIT press. |
| [GH14] | Faust, GG. and Hall, IM. (2014). "SAMBLASTER: fast duplicate marking and structural variant read extraction," Bioinformatics, 30(17):2503–2505 |
| [GNT+10] | Goecks, J., Nekrutenko, A. and Taylor, J. (2010). "Galaxy: a comprehensive approach for supporting accessible, reproducible, and transparent computational research in the life sciences." Genome Biol 11(8): R86. |
| [KKLS17] | Kanwal, S., Khan, F. Z., Lonie, A. and Sinnott, R. O. (2017). "Investigating reproducibility and tracking provenance - A genomic workflow case study." BMC Bioinformatics 18(1): 337. |
| [KR12] | Köster, J. and Rahmann, S. (2012). "Snakemake—a scalable bioinformatics workflow engine." Bioinformatics 28(19): 2520-2522. |
| [Lei17] | Leipzig, J. (2017). "A review of bioinformatic pipeline frameworks." Brief Bioinform 18(3): 530-536. |
| [LHW+09] | Li, H., Handsaker, B., Wysoker, A., Fennell, T., Ruan, J., Homer, N., et al. (2009). "The sequence alignment/map format and SAMtools". Bioinformatics, 25(16), 2078-2079. |
| [Li13] | Li, H. (2013). "Aligning sequence reads, clone sequences and assembly contigs with BWA-MEM". arXiv preprint arXiv:1303.3997. |
| [LPVM15] | Liu, J., Pacitti, E., Valduriez, P. and Mattoso, M. (2015). "A Survey of Data-Intensive Scientific Workflow Management." Journal of Grid Computing *13*(4), 457-493. |
| [MDBL16] | McDougal, R. A., Bulanova, A. S., and Lytton, W. W. (2016). "Reproducibility in computational neuroscience models and simulations". IEEE Transactions on Biomedical Engineering, 63(10):2021-2035. |
| [MLL+16] | Muir, P., Li, S., Lou, S., Wang, D., Spakowicz, D. J., Salichos, L., Zhang, J., Weinstock, G. M., Isaacs, F., Rozowsky, J., et al. (2016). "The real cost of sequencing: scaling computation to keep pace with data generation." Genome Biol 17(1): 53. |
| [MWHW16] | Missier, P., Woodman, S., Hiden, H., and Watson, P. (2016). "Provenance and data differencing for workflow reproducibility analysis". Concurrency and Computation: Practice and Experience, 28(4):995-1015. |
| [NT14] | Nekrutenko, A. and Taylor, J. (2012). "Next-generation sequencing data interpretation: enhancing reproducibility and accessibility." Nat Rev Genet 13(9): 667-72. |
| [OST+19] | Ossyra, J., Sedova, A., Tharrington, A., Noé, F., Clementi, C., and Smith, J. C. (2019). "Porting adaptive ensemble molecular dynamics workflows to the summit supercomputer". Int. Conf. on High Performance Computing. |
| [PFV+15] | Pradal, C., Fournier, C., Valduriez, P., and Cohen-Boulakia, S. (2015). "OpenAlea: scientific workflows combining data analysis and simulation". Int. Conf. on Scientific and Statistical Database Management. |
| [PFvdS17] | Plöthner, M., Frank, M., and von der Schulenburg, J. M. G. (2017). "Cost analysis of whole genome sequencing in German clinical practice". The European Journal of Health Economics, 18(5):623-633. |
| [RDM+12] | Roy, A., Diao, Y., Mauceli, E., Shen , Y. and Wu, B.-L. (2012). "Massive Genomic Data Processing and Deep Analysis". Int. Conf. on Very Large Databases Istanbul, Turkey. |
| [SBR+15] | Spjuth, O., Bongcam-Rudloff, E., Hernandez, G. C., Forer, L., Giovacchini, M., Guimera, R. V., Kallio, A., Korpelainen, E., Kandula, M. M., Krachunov, M., et al. (2015). "Experiences with workflows for automating data-intensive bioinformatics." Biol Direct 10: 43. |
| [SBTW18] | Schwarze, K., Buchanan, J., Taylor, J.C. and Wordsworth, S., 2018. Are whole-exome and whole-genome sequencing approaches cost-effective? A systematic review of the literature. Genetics in Medicine, 20(10):1122-1130. |
| [SCFB15] | Shringarpure, Suyash S., Andrew Carroll, M. Francisco, and Carlos D. Bustamante. "Inexpensive and highly reproducible cloud-based variant calling of 2,535 human genomes." PloS one 10(6). |
| [SCN+15] | Segal, O., Colangelo, P., Nasiri, N., Qian, Z., and Margala, M. (2015). "Sparkcl: A unified programming framework for accelerators on heterogeneous clusters". arXiv preprint arXiv:1505.01120. |
| [SL10] | Simmhan, Y., and Ramakrishnan, L. (1010). "Comparison of resource platform selection approaches for scientific workflows." Int. Symp. on High Performance Distributed Computing. |
| [SSM18] | Stodden, V., Seiler, J., and Ma, Z. (2019). "An empirical analysis of journal policy effectiveness for computational reproducibility." Proceedings of the National Academy of Sciences 115(11):2584-2589. |

[SSVS15]    Siretskiy, A., Sundqvist, T., Voznesenskiy, M. and Spjuth, O. (2015). "A quantitative assessment of the hadoop framework for analyzing massively parallel dna sequencing data." Gigascience 4(1).

[TBG+12]    Tröger, P., Brobst, R., Gruber, D., Mamonski, M., and Templeton, D. (2012). "Distributed resource management application API Version 2 (DRMAA)". Technical report, Open Grid Forum.

[WGL18]    Witt, C., Bux, M., Gusew, W. and Leser, U. (2018). "Predictive Performance Modeling in Distributed Computing using Black-Box Monitoring and Machine Learning". arXiv:1805.11877.

[WZX+16]    Wu, D., Zhu, L., Xu, X., Sakr, S., Sun, D., and Lu, Q. (2016). "Building pipelines for heterogeneous execution environments for big data processing". IEEE Software, 33(2), 60-67.

[ZCD+12]    Zaharia, M., Chowdhury, M., Das, T., Dave, A., Ma, J., McCauley, M., Franklin, M. J., Shenker, S. and Stoica, I. (2012). "Resilient distributed datasets: a fault-tolerant abstraction for in-memory cluster computing". USENIX conference on Networked Systems Design and Implementation. San Jose, USA.

[ZLJ+13]    Zou, Q., Li, X. B., Jiang, W. R., Lin, Z. Y., Li, G. L. and Chen, K. (2013). "Survey of MapReduce frame operation in bioinformatics." Brief Bioinform, 15(4), 637-647.

# Supplementary Material

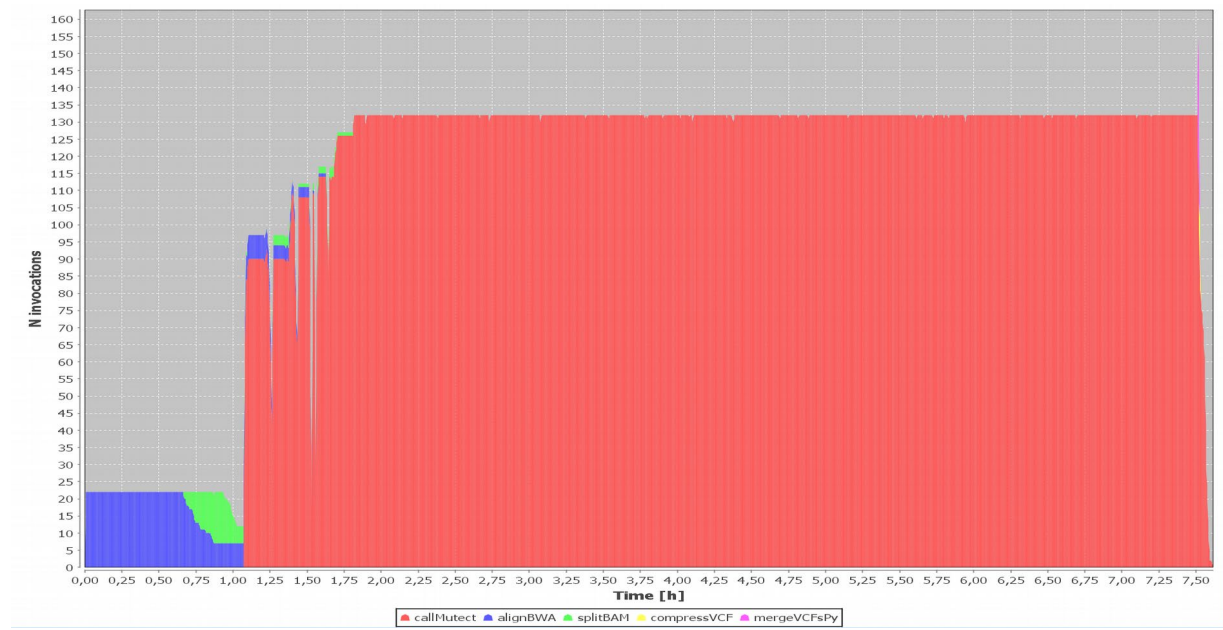

Figure SM1: Task invocations over time on YC with 467 MuTect tasks per alignment pair. The alignment process (blue) is followed by splitting the alignment file (green) and subsequent variant identification (red).

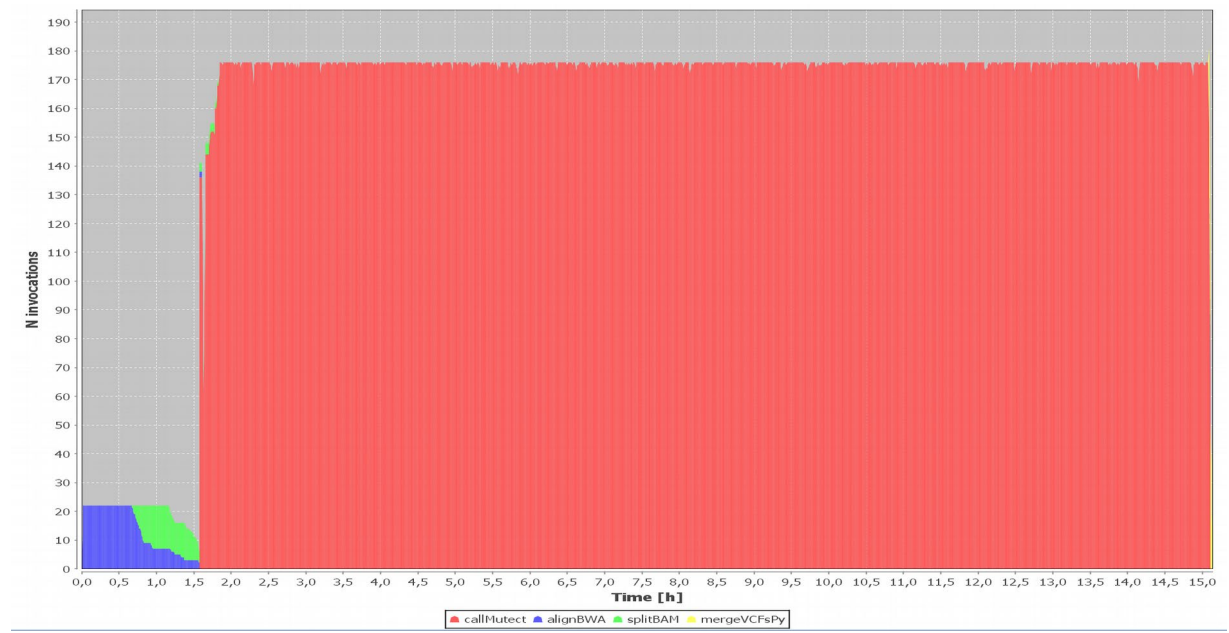

Figure SM2: Task invocations over time on YC with 2863 MuTect tasks per alignment pair. See Figure SM1 for color codes.